# Nanoscale

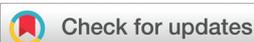

ROYAL SOCIETY OF CHEMISTRY

Check for updates

## Tuning nanowire lasers *via* hybridization with two-dimensional materials†

Edwin Eobaldt, [ID] [a] Francesco Vitale,[a] Maximilian Zapf, [ID] [a] Margarita Lapteva,[a] Tarlan Hamzayev,[a] Ziyang Gan,[b] Emad Najafidehaghani,[b] Christof Neumann, [ID] [b] Antony George, [ID] [b] Andrey Turchanin, [ID] [b,c] Giancarlo Soavi [ID] *[a,c] and Carsten Ronning [ID] *[a,c]

Mixed-dimensional hybrid structures have recently gained increasing attention as promising building blocks for novel electronic and optoelectronic devices. In this context, hybridization of semiconductor nanowires with two-dimensional materials could offer new ways to control and modulate lasing at the nanoscale. In this work, we deterministically fabricate hybrid mixed-dimensional heterostructures composed of ZnO nanowires and MoS₂ monolayers with micrometer control over their relative position. First, we show that our deterministic fabrication method does not degrade the optical properties of the ZnO nanowires. Second, we demonstrate that the lasing wavelength of ZnO nanowires can be tuned by several nanometers by hybridization with CVD-grown MoS₂ monolayers. We assign this spectral shift of the lasing modes to an efficient carrier transfer at the heterointerface and the subsequent increase of the optical band gap in ZnO (Moss–Burstein effect).

## Introduction

The increasing technological demand for ever-smaller components and the current limit of integrated circuits has pro-

[a]Institute of Solid State Physics, Friedrich Schiller University Jena, 07743 Jena, Germany. E-mail: giancarlo.soavi@uni-jena.de, carsten.ronning@uni-jena.de
[b]Institute of Physical Chemistry, Friedrich Schiller University Jena, 07743 Jena, Germany
[c]Abbe Center of Photonics, Friedrich Schiller University Jena, 07745 Jena, Germany
†Electronic supplementary information (ESI) available. See DOI: 10.1039/d1nr07931j

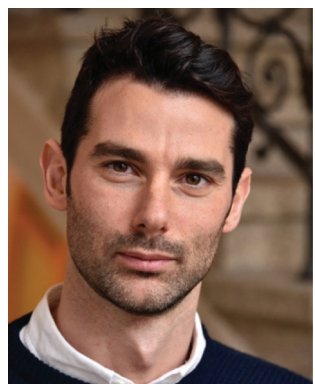

**Giancarlo Soavi**

*Giancarlo Soavi obtained a PhD in Physics from Politecnico di Milano (Italy) in 2015 and subsequently worked as a Research Associate at the Cambridge Graphene Centre (UK). From 2019 he is a tenure track Professor at the Friedrich Schiller University of Jena (Germany). His research interests focus on ultra-fast spectroscopy and nonlinear optics in quantum confined systems, including graphene and related layered materials.*

moted intense research towards nanoscale electronic and optoelectronic devices. Among those, hybrid systems obtained by combining zero- (0D), one- (1D) and two-dimensional (2D) materials are particularly promising. Enabled by the efficient light absorption and high surface-to-volume ratios of nanoscale quantum confined materials, such hybrid systems enable the effective separation of excited carries at the heterointerface, offering new opportunities for high-performance solar cells and photodetectors.[1,2] Beyond that, hybrid nanostructures are also promising candidates for novel spintronic devices,[3,4] nanoscale lasers[5–8] and gas sensors.[9–11]

Within the family of 1D materials, semiconductor nanowires (NWs) are of particular scientific and technological interest. In recent years, they have attracted great scientific attention due to their excellent material quality and remarkable waveguiding properties (*i.e.*, 1D light confinement). One of the most appealing properties of semiconductor NWs is their capability to lase without the need of external cavities: they provide optical gain under sufficiently high pumping and act as efficient optical resonators due to their morphology and high refractive index contrast at the end facets.[12] Following early observations of lasing in semiconductor NWs,[13] scientific research has increasingly focused on exploring its spectral and temporal characteristics.[14] In the last decade, wavelength tunability[14] and continuous wave operation[15] in NW lasers have been demonstrated. Additionally, semiconductor NWs can be fabricated with precise control over their shape and dimension both in bottom-up and top-down approaches,[16,17] thus

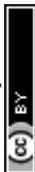

offering a viable platform for future scalable technological applications.

In parallel, 2D materials appear as ideal candidates for the engineering of hybrid heterostructures, also thanks to their largely tunable electrical[18,19] and optical properties.[20–22] Previous studies on 1D–2D heterostructures formed by semiconductor NWs and layered materials focused mainly on tuning the non-linear optical properties of layered materials,[23] enhancing plasmonic lasing on metallic substrates,[5,6] and realizing high-responsivity photodetectors[24,25] and high-sensitivity gas sensors.[26–28] However, to the best of our knowledge, to date no comprehensive study on the lasing properties of hybrid NW and layered materials heterostructures was published. In addition, most of such hybrid devices were fabricated either by epitaxial growth of NWs on 2D templates,[29–32] or by drop-casting a NW dispersion on a substrate and subsequent placement of the layered material on top of it, typically with a stage-assisted[33] polydimethylsiloxane (PDMS) transfer or similar lift-off processes.[34–37] Only very recently, such a PDMS-based transfer approach has been successfully used also for NWs.[34,38]

In this work, we present a systematic study of the stage-assisted transfer technique for NWs that are placed on top of chemical vapor deposition (CVD)-grown $MoS_2$ crystals with micrometer precision. In particular, we focus on the photonic lasing characteristics of $MoS_2$ monolayer-hybridized ZnO NWs and we demonstrate that the lasing wavelength of ZnO shifts by several nanometers due to interfacial charge transfer and the Moss–Burstein effect. As a consequence of their high optical gain supplied by the formation of an electron hole plasma, ZnO NWs provide robust ultraviolet lasing emission at room temperature.[39–42] On the other hand, $MoS_2$ MLs emit light in the visible range ($E_g \approx 1.8$ eV[43]) which, in turn, does not spectrally overlap with the NW lasing emission. Additionally, the lattice constants of these materials match almost perfectly,[42] making them ideal candidates for hybrid 1D–2D heterostructures.

## Experimental

### ZnO nanowire and $MoS_2$ monolayer synthesis

The ZnO NWs used in this study were fabricated by a vapor transport technique exploiting the vapor–liquid–solid (VLS) mechanism[44] inside a horizontal three-zone tube furnace.[45] The source material, consisting of a mixture of pure ZnO powder and graphite (mass ratio 7 : 1), was placed in the center of the tube furnace and sublimated at 1050 °C. A carrier gas mixture containing Ar and $O_2$ (10 sccm each) transported the source vapor towards Au-coated ($\approx$10 nm thin film) silicon substrates, which were kept at 1000 °C in the end zone of the furnace. The growth time was 60 min at a pressure of 7 mbar.

$MoS_2$ MLs with high optical quality were fabricated by CVD inside a two-zone split furnace.[46,47] Due to their crystal symmetry, CVD-grown $MoS_2$ monolayers tend to grow in equilateral triangular flakes. Detailed information on the setup and the growth parameters can be found in ref. 46.

### Deterministic nanowire transfer

We used a stage-assisted technique[33] to deterministically imprint ZnO NWs on a target substrate by using PDMS stamps. A sketch of the transfer stage setup is provided in Fig. S1 of the ESI.†

As a first step, as-grown NWs were dry-imprinted onto clean $SiO_2$/Si substrates ($t_{ox} \approx 300$ nm). PDMS pieces (typically about 1 $mm^2$ large), fixed below a microscope cover glass, were used to stamp NWs from the imprinted to a target substrate. By using a commercial transfer stage setup (HQ graphene) with long working distance objectives (10×, 20×, 50× magnification), a selective and high-precision transfer of single NWs was achieved. The position and the orientation of the transferred NWs could be adjusted with a precision of 1 μm and 2°, respectively.

### Micro-photoluminescence (μPL) measurements

For the optical characterization of single NWs, a frequency-tripled Nd:YAG laser ($\lambda_{ex} = 355$ nm, $f_{rep} = 100$ Hz, $t_{pulse} = 7$ ns) was defocused to spot diameters of 20–30 μm with a 50× NUV objective (NA = 0.43). Neutral density filters in the beamline allowed for an attenuation of the laser intensity by several orders of magnitude, while a silicon photodiode was used for *in situ* power measurements. The NW emission was collected perpendicular to the substrate plane by the same objective and focused into a 500 mm monochromator (Princeton Instruments SP-2500i). A 1200 grooves per mm grating ($\sim$0.1 nm resolution) allowed for the dispersion of the emitted light which was subsequently detected by a liquid-nitrogen-cooled, front-illuminated CCD camera. All optical measurements were performed at room temperature. Each measurement consisted of a power series, in which several luminescence spectra were recorded at different excitation powers. For this purpose, the laser spot of the pulsed Nd:YAG laser was focused down to diameters slightly larger than the NW length (typically 20–25 μm) and we collected the luminescence emitted along the whole NW.

### Photoelectron spectroscopy

X-ray photoelectron spectroscopy (XPS) and UV photoelectron spectroscopy (UPS) were performed in an ultra-high-vacuum ($<2 \times 10^{-10}$ mbar) Multiprobe System (Scienta Omicron) using a monochromatized X-ray source (Al $K_\alpha$, 1486.7 eV) and a gas discharge He I vacuum UV light source (21.22 eV), respectively. The electron analyzer (Argus CU) had a spectral resolution of 0.6 eV (XPS) and 0.1 eV (UPS). The pass energy amounted to 50 eV for the survey XP spectra, 30 eV for high-resolution XP spectra and 5.5 eV for UP spectra. The UPS measurements were recorded using a negative bias voltage of 10 V, which is corrected in the presented spectra (see Fig. S8 of the ESI†). The XP spectra (see Fig. S7†) were fitted using Voigt functions (30 : 70) after background subtraction. The spectra were peak shift-calibrated using the Zn $2p_{3/2}$ peak at a binding energy of 1022.1 eV (ZnO). For a quantitative estimate, we used the relative sensitivity factors 5.62 (Mo $3d_{5/2}$) and 1.11 (S $2p_{3/2}$).

# Results and discussion

## Transfer-induced changes of ZnO nanowire lasing

In order to test our PDMS-based and stage-assisted transfer technique, we transferred more than 80 individual NWs from dry-imprinted SiO$_2$/Si substrates to similar clean target substrates while tuning the most important process parameters. The transfer efficiency primarily depended on the peel velocity and the temperature of the stamp, which is related to the viscoelasticity of PDMS. As a result, higher temperatures and low strains led to a significantly lower adhesion of the PDMS stamp to the nanostructure, and *vice-versa* for the opposite case.[48] Hence, NWs were attached to the PDMS at room temperature with high peel velocities of about 0.5–1 cm s$^{-1}$. The attachment efficiency (≈4%) was calculated from the average number of trials necessary to attach the NW to the stamp. On the contrary, the NW was released from the stamp to the target substrate by heating the PDMS and applying low peel velocities of about 20–100 μm s$^{-1}$. At 80 °C, a maximum release efficiency of ≈85% was found, and it did not significantly rise for higher temperatures.

Next, the lasing characteristics of eight NWs were analyzed under similar experimental conditions before and after the transfer. As a reference, a control group of seven NWs was measured twice on the same substrate. For the sake of comparability and to rule out possible storage-related effects,[49] the repeated measurements for all NWs were carried out three days after the first ones. All NWs originated from the same growth batch, and they featured lengths between 10–22 μm and diameters between 220–350 nm.

In Fig. 1a, the power series of the first and the second measurements are exemplarily compared for the same transferred NW. At low excitation powers (blue graph), the NW emitted in a broader spectral region around 378 nm, which is associated with the spontaneous recombination of excitonic states close to the conduction band, *i.e.* the so-called near band edge emission (NBE). Upon increasing the excitation power, sharp and equidistant modes appeared on top of the spontaneous emission background at around 380 nm, which is related to the amplified spontaneous emission (ASE). Upon a further increase of the excitation power, we observed the build-up of Fabry–Pérot modes, indicating the lasing action in the NW.

Moreover, we determined the input-output characteristics of the NWs before and after the transfer, as shown in Fig. 1b. The experimental values for the integrated PL intensity were fitted with a multi-mode laser model[50] which provided an estimation for the lasing threshold intensity $I_{th}$. The same analyses were carried out for the transferred NWs and the control group (Fig. 1c). Remarkably, all the transferred NWs retained their ability to lase. Thus, a highly detrimental effect induced by the transfer can be excluded. Fig. 1c further illustrates the relative threshold change of each NW (transferred and control groups) after the second measurement: both groups show a similar distribution with nearly equal mean values (indicated by the dashed lines) and standard deviations (indicated by the

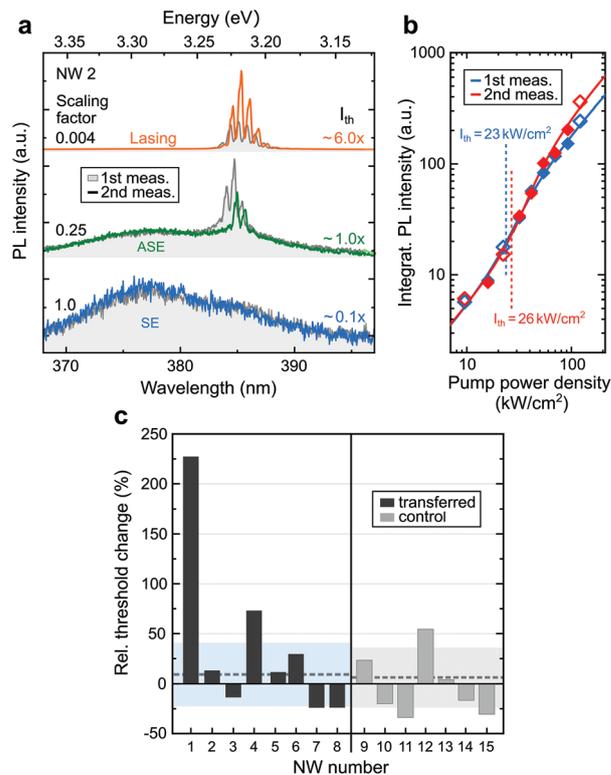

**Fig. 1** (a) Photoluminescence spectra of a single nanowire below (blue), at (green), and above (orange) the lasing threshold. All spectra were re-measured after the transfer (solid, colored lines) and are plotted together with the initial measurement (grey segments). (b) Pump power density dependence of the integrated nanowire emission for the first (blue squares) and the second (red squares) measurements of the same nanowire. The spectra are shown at the same multiples of the respective threshold. The empty squares correspond to the spectra shown in (a). The solid lines represent a fit to the data using a multimodal fit model[34] from which the lasing threshold intensity $I_{th}$ was estimated. The integrated PL intensity was obtained by integrating each spectrum over a specific wavelength region around the modal features. (c) Relative change of the lasing threshold after the second measurement shown for the transferred NWs and the control group. The colored segments illustrate the standard deviations of both groups. The dashed lines depict the corresponding mean values. Excluding NW1, both groups exhibit a similar distribution.

colored segments). The latter amounted to 30% for the control group and 34% for the transfer group. Further information on the absolute threshold values for both measurements are given in Fig. S2 of the ESI.† The first NW, which exhibited a threshold change that was thrice the initial value, was considered as an outlier and, therefore, excluded from the calculations. Apart from this, there was no evidence for a transfer-related influence on the NW lasing threshold. The changes of the lasing thresholds for both groups can rather be attributed to small intrinsic deviations of the experimental conditions between the PL measurements. For instance, it is reasonable to assume an uneven excitation power distribution along the NW – mostly due to hotspots and inhomogeneities in the spatial profile of the beam spot, which can affect the spatial gain distribution and, thus, the lasing threshold of the device.

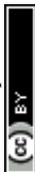

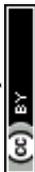

Subsequently, we investigated the effect of the PDMS assisted transfer on the modal structure of the NW lasing. To this end, we compared the lasing spectra of the same NW before and after the transfer, as exemplarily illustrated in Fig. 2a. The normalized spectra were recorded at the same multiple of the respective thresholds, normalized by their maximum intensity, and the spontaneous emission background was subtracted. We observed slight differences in the individual mode intensity as well as in the width of the spectral region in which the modes appeared. Furthermore, each individual mode was slightly shifted by a nearly constant value, as exemplarily indicated by the dashed lines. The upper diagram in Fig. 2b depicts this mode shift for each NW. Here, the mean values are almost the same the transfer group, however, showed a moderately higher fluctuation. In order to quantify the spectral width change of the mode region, we used the full width at half maximum (FWHM) of a Gaussian fit function. The mean value of the relative width change was again comparable for both NW groups, as illustrated in the lower diagram of Fig. 2b. Again, the transferred NWs showed a slightly higher fluctuation around the mean value. However, this alternation of the values cannot be explained by transfer-related effects, such as surface contamination, a change of the NW geometry or strain induced by the transfer process. Instead, slightly modified excitation conditions are more likely responsible for such changes: since the effective refractive index $n$ of the NW depends on the excitation intensity,[51] the individual mode positions will shift to match the constructive interference condition: $\lambda \sim n$. Analogously, the gain profile effectively changes with intensity, leading to a different mode support and, therefore, might explain the different intensities of the individual modes as well as the width change of the spectral mode region.

Finally, two NWs were transferred a second time in order to confirm any potential effects induced by the first transfer attempt. The corresponding threshold and mode structure diagrams are presented in Fig. S3 of the ESI.† Even after a repeated transfer, no anomalies between the two groups were observed. This result supports the previous finding that the PDMS transfer does not (or only barely) affect the NW lasing characteristics, in agreement with ref. 38.

## Lasing of MoS₂ monolayer-hybridized ZnO nanowires

The optimized stage-assisted PDMS transfer method allowed us to place individual NWs on CVD-grown $MoS_2$ monolayers with high precision in terms of positioning and orientation, offering mixed-dimensional hybrid structures, as shown in Fig. 3a–c. In the following, the hybridized NWs are mainly characterized by their relative overlap, which refers to the fraction of the NW lying on the monolayer flake.

To examine the effect of the underlying MoS2 ML on the lasing properties of the ZnO NW, we studied the changes in the lasing thresholds and lasing modes as a function of the relative overlap. For this purpose, we fabricated and measured a total of six different hybridized NWs. The optical images in Fig. 3a–c show a selection of $MoS_2$ monolayer-hybridized ZnO NWs with relative overlaps ranging from 9% to 100%. The investigated NWs had lengths between 16–24 μm and diameters from 170–240 nm, while the CVD-grown monolayer flakes possessed side lengths between 15 and 30 μm. Additional atomic force microscopy (Fig. S4 of the ESI†) and Raman (Fig. S5†) measurements also confirmed the monolayer character of the $MoS_2$ crystals. The lasing measurements were again performed before and after the transfer using the same PL setup as in the transfer experiments described above. However, in this case, only the emission from the hybridized NW end facets was collected. In Fig. 3d, the relative lasing threshold change of the hybridized NWs is plotted against the relative overlap. As known from the previous experiment, the relative threshold changes of the PDMS-transferred NWs statistically fluctuated with a standard deviation of about 34%.

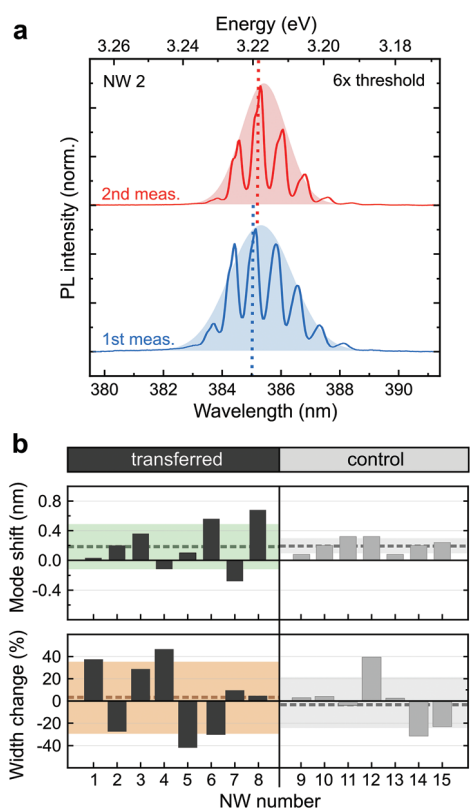

**Fig. 2** (a) Comparison of the modal structure for the first and second measurement of the same NW. The spectra were normalized to their corresponding maximum, and the respective spontaneous emission background was subtracted as baseline. A Gaussian function was used to generate a symmetric mode envelope for both graphs. Besides differences in the mode intensities, the individual mode positions slightly shift by the same value, as depicted by the dashed lines. The most noticeable feature is the different width of the mode profiles, whose relative change can be described by the FWHM of the corresponding mode envelope. (b) Systematic analysis of the modal changes for both NW groups. Top: spectral shift of the central lasing mode. Bottom: relative change of the envelope width. The standard deviation and mean values of both groups are illustrated in analogy to Fig. 1c. No evident deviations are present between the two groups, even though the values of the transferred NWs fluctuate more.

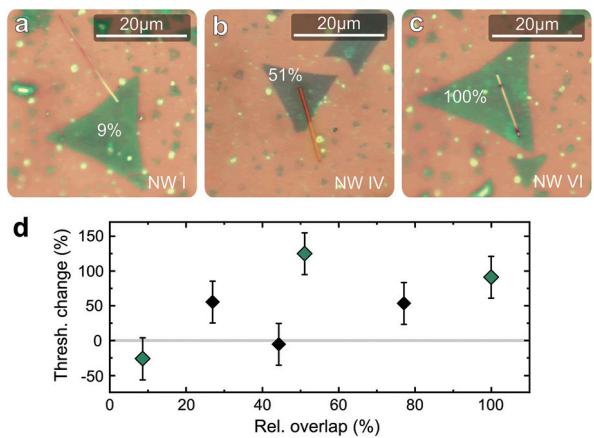

**Fig. 3** (a)–(c) Optical microscope images of ZnO NWs-MoS₂ monolayer hybrids with relative overlaps of 9%, 51%, and 100% respectively. (d) Relative lasing threshold change of the MoS₂-hybridized ZnO NWs plotted against their relative overlap. The green squares correspond to the hybrid structures in (a)–(c). The error bars of 34% represent the statistical threshold uncertainty of the PDMS-transferred NWs (see Fig. 1c). For higher values of relative overlap to the monolayer flake, the lasing threshold of the NWs significantly increased.

The latter can be considered as an estimation for the measurement uncertainty and is, therefore, represented by the error bars of the data points presented here. The lasing thresholds of the hybridized NWs were significantly increased for higher values of the overlap, even exceeding relative changes of more than 150%. However, no clear systematic dependence on the relative overlap can be assessed. Hence, we could not verify any unambiguous effect caused by the presence of the monolayer on the lasing threshold. However, since the higher threshold may originate from increased losses in the hybridized part of the NW due to a modification of the dielectric environment.

Subsequently, we studied the modal changes of the NW emission after hybridization with MoS₂. Fig. 4a–c display the lasing spectra obtained under the same experimental conditions as for the data shown in Fig. 2b: however, in this case, both measurement series (accomplished before and after the transfer) were recorded at the same absolute power as well as at the same multiple of the lasing threshold. A strong blue shift of the NW lasing modes is clearly seen to occur for higher values of the relative overlap. On the contrary, other mode characteristics, such as the change of the mode envelope width or the individual mode shift, did not show any anomalies with respect to the transfer experiment as reference (see Fig. 2b). The spectral mode shift was further quantified by examining the shift of the Gaussian mode envelope maximum (marked by the dashed lines) for each hybridized NW, respectively (see Fig. 4c). Clearly, the modal blue shift is less than 1 nm for small overlaps, and it increases almost monotonously for increasing relative overlaps up to a maximum value of ≈3.5 nm for a full (100%) overlap. All the investigated NWs had similar lengths and diameters and thus allow for a reasonable qualitative comparison between the different wires.

As a possible explanation for the observed blue shift of the NW lasing modes as a function of the relative overlap with the MoS₂ flake we propose the Moss–Burstein (MB) effect, due to an electron transfer from the MoS₂ ML to the ZnO NW. The MB effect was already proposed as a potential origin of the modal blue shift of CdS NW lasers on Al substrates.[52] In Fig. 5a, we schematically show the expected band alignment, which supports the proposed carrier transfer. At the interface between the two structures a type II heterojunction forms where the conduction band of ZnO is slightly below that of MoS₂. Upon optical excitation, charge separation and transfer will occur, with excited electrons drifting from the MoS₂ conduction band to the that of ZnO. Analogously, the holes will accumulate in the valence band of MoS₂. Due to the lower

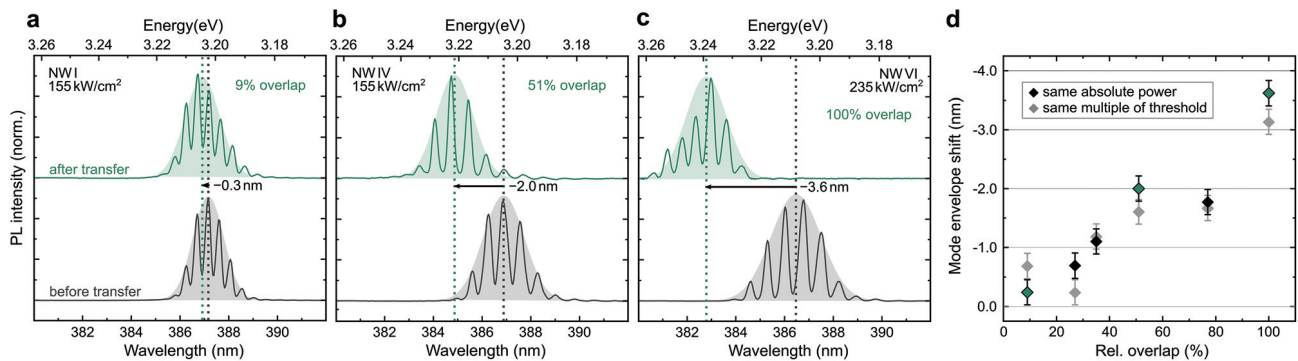

**Fig. 4** (a)–(c) Comparison of the normalized lasing spectra of the hybridized NWs of Fig. 3a–c before and after the transfer measured at the same excitation power. All spectra were obtained similarly to the measurements in Fig. 2a. The dashed lines indicate the respective maxima of the Gaussian mode envelopes. An increase of the relative overlap is associated with a larger blue shift of the envelope maxima up to 3.5 nm for a full overlap. (d) Spectral shift of the lasing mode envelope plotted against the relative overlap of the MoS₂-hybridized NWs obtained for the same absolute laser power and the same multiple of the threshold (grey squares). The green squares correspond to the hybrid structures in Fig. 3a–c. The error bars are estimated to be 0.2 nm and originate from statistical fluctuations in fitting the envelope to the spectra and slightly altered measurement conditions. The lasing modes of the NWs experienced a considerable blue shift that is positively correlated to their relative overlap with the monolayer flake.

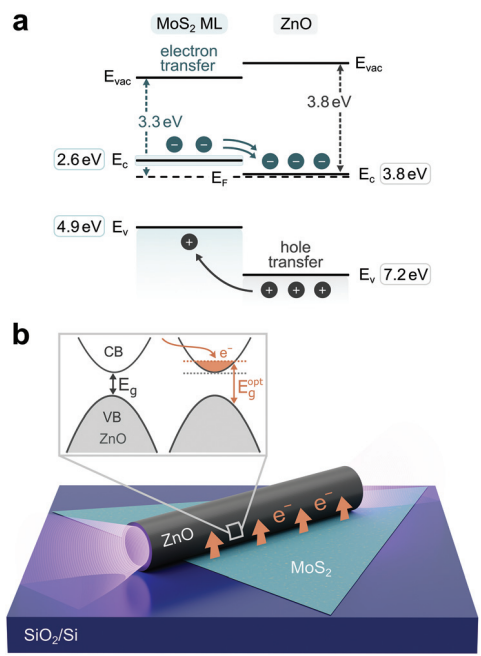

**Fig. 5** (a) Calculated band alignment of the investigated hybrid structures. The work functions and valence band levels were measured with XPS/UPS measurements (see Fig. S8†) and are given with respect to the vacuum level. For the band gap energy of ZnO a value of 3.4 eV was taken from the literature.[42] The band gap energy of the $MoS_2$ ML was estimated by considering its optical band gap of 1.8 eV (ref. 43) and its large exciton binding energy, which is in the range of 200–900 meV.[57–59] In sum, a type II heterojunction forms at the interface between the two materials and leads to an effective carrier separation upon optical excitation. Due to their higher mobility, the electron transfer towards the NW is the dominant process (b) ZnO NW partially lying on a CVD-grown $MoS_2$ monolayer on top of a $SiO_2/Si$ substrate. Inset: the carrier transfer at the NW/monolayer interface leads to an electron excess in the conduction band of the NW which widens its optical band gap (Moss–Burstein effect). As a result, recombining electrons emit radiation with a considerably higher energy.

mobility of the holes, electron transfer represents the dominant process here. Hence, the ZnO conduction band is partially filled with excess electrons which results in an effective increase of the optical band gap (see Fig. 5b) and, as a consequence, in a considerably higher energy of the generated lasing modes.

We note that, while there are publications that point towards the previously discussed type I band alignment and electron transfer at the $ZnO/MoS_2$ interface,[27,28] other works predict a type II alignment with opposite (hole) carrier transfer[53,54] or even a type I alignment.[55] As the band alignment is also related to a close interplay of diverse factors, such as the nanostructure fabrication, the underlying substrate, the presence of surface defects *etc.*, we additionally conducted the following experiment: $MoS_2$ flakes, grown with the identical parameters described above, were transferred on top of ZnO single crystalline substrates (with ⟨0001⟩ orientation) by using a PMMA-based transfer method.[46,56] This sample, along with a pristine ZnO sample, were investigated by XPS/UPS measure-ments without any further cleaning procedure in order to mimic the same surface conditions of our lasing experiments. The XP/UP spectra as well as their analysis are presented in the ESI (Fig. S7 and S8†) and clearly confirm a type II band alignment of $MoS_2$ and ZnO. The determined work functions and valence band energy levels have been added to Fig. 5a. The corresponding band gap energies of 1.8 eV for $MoS_2$ MLs and 3.4 eV for ZnO were taken from the literature.[42,43] Since a value of 1.8 eV reflects the optical band gap of $MoS_2$ MLs, its electronic band gap was calculated by adding the exciton binding energy, which is about 200–900 meV,[57–59] depending on the dielectric constant of the substrate.[60] However, the exact binding energy is difficult to access for our geometry. Nonetheless, even considering low values for the binding energy uncertainty, the hybrid heterostructure retains its type II band alignment at the heterointerface.

## Conclusions

In this work, we presented a deterministic approach to fabricate 1D–2D heterostructures composed of ZnO NWs and CVD-grown $MoS_2$ MLs. To this end, we specifically optimized a stage-assisted PDMS transfer technique for nanowires, allowing for a precise control over their position and orientation. Comparative PL measurements revealed that the lasing properties of the ZnO were not (or only barely) affected by this transfer process.

In addition, we have shown that the hybridization of the NWs with $MoS_2$ produces a significant spectral blue shift of the lasing modes up to several nanometers. Supported by XPS and UPS measurements, this blue shift can be assigned to the formation of a type II band alignment at the heterointerface, leading to an effective separation and transfer of electrons from $MoS_2$ to ZnO. This band alignment results in an increase of the ZnO optical band gap (Moss–Burstein effect) induced by the carrier injection into the NW conduction band.

Our optimized NW transfer approach distinguishes itself by its simplicity to deterministically fabricate NW-monolayer hybrids without affecting the nanowire lasing properties. This methodology will pave the way towards the fabrication of more complex architectures, such as NW lasers fully encapsulated between layered materials for the engineering of the refractive index dispersion, as well as the realization of all-optically and electrically tunable nanoscale lasers by carefully controlling and engineering the charge transfer processes in these hybrid structures. Our results thus provide a new platform for advanced nanoscale photonic and optoelectronic devices.

## Author contributions

E. E., C. R. and G. S. conceived the experiments. E. E. prepared the samples; E. E., F. V. and M. Z. performed the PL experiments. E. N. Z. G., C. N., A. G. and A. T. prepared the CVD-grown $MoS_2$ monolayers and $MoS_2/ZnO$ single crystal heterostructures, characterized these samples with AFM, XPS/UPS and Raman spectroscopy and analyzed the results. T. H. and M. L. built the PDMS transfer stage setup and helped with the sample preparation. The results were discussed and interpreted by all authors. E. E., C. R. and G. S. wrote the manuscript with feedback from all co-authors.

## Conflicts of interest

There are no conflicts of interest to declare.

## Acknowledgements

This work was supported by the European Union's Horizon 2020 Research and Innovation Program under Grant Agreement GrapheneCore3 881603 (G. S.). The authors acknowledge the German Research Foundation DFG (CRC 1375 NOA project numbers C5 (C. R.), B2 (A. T.) and B5 (G. S.)) and the Daimler und Benz Foundation (G. S.) for financial support.